\documentstyle[aps,preprint,prb,epsf]{revtex}

\begin{document}

\title{For A Lecture on Scientific Meteorology within Statistical 
(''Pure'') Physics Concepts }

\author{  M. Ausloos }

\address{ SUPRATECS\footnote{SUPRATECS = Services Universitaires Pour la
Recherche et les Applications Technologiques de mat\'eriaux Electroc\'eramiques,
Composites et Supraconducteurs}, Institute of Physics, B5, \\University of Li$\grave
e$ge, B-4000 Li$\grave e$ge, Belgium}

\date{\today}


\draft

\maketitle

\abstract{Various aspects of modern statistical physics and meteorology can be
tied together. Critical comments have to be made. However, the historical importance of the University of Wroclaw in
the field of meteorology should be first pointed out. Next, some basic difference
about time and space scales between meteorology and climatology can be outlined.
The nature and role of clouds both from a geometric and thermal point of view are
recalled. Recent studies of  scaling laws for atmospheric variables are
mentioned, like studies on cirrus ice content, brightness temperature, liquid
water path fluctuations, cloud base height fluctuations, ....  Technical time
series analysis approaches based on modern statistical physics considerations are
outlined.}

\vskip 1cm

\section{ INTRODUCTION and FOREWORD}

This  contribution to the 18th Max Born Symposium Proceedings, cannot be seen
as an extensive review of the connection between meteorology and various aspects
of modern statistical physics. Space and time (and weather) limit its content.
Much of what is found here can rather be considered to result from a biased view
point or attitude  and limited understanding of the author frustrated because the
developments of what he thought was a science (meteorology) turns out to be
unsatisfactory to him and sometimes misleading. It seems that other approaches
might be thought of. New implementations carried forward. Some are surely made
and understood by meteorologists but are not easily available in    usual physics
literature. Thus the lines below may  be rather addressed to physicists. The
paper will be satisfactory if  it attracts work toward a huge field of interest
with many many publications still with many unanswered questions. As an immediate
warning it is emphasized that deep  corrections to standard models or actual
findings can NOT be found here nor are even suggested. Only to be found is a set
of basic considerations and reflections expecting to give lines of various
investigations, and hope for some scientific aspects of meteorology in the spirit
of modern statistical physics ideas.

A historical point is in order. The author came into this subject starting from
previous work in econophysics, when he observed that some "weather derivatives"
were in use, and some sort of game initiated by the Frankfurt Deutsche
B\"{o}rse\cite{xelsius} in order to attract customers which could predict the
temperature in various cities within a certain lapse of time, and win some prize
thereafter\footnote{A common measure of temperature has arisen from the market :
the  degree-day. The Heating Degree Day is a (loose) measure of how much heating
is needed. For a given day, HDD is equal to the larger of Max [0, 18$^{\circ}$C -
daily average temperature]. The Cooling Degree Day (CDD) is a (loose) measure of
how much cooling is needed:  CDD = Max[0, daily average temperature -
18$^{\circ}$C].   This notion seems to be a measure that the energy suppliers
could use to hedge their supply in adverse temperature conditions.}. This subject
therefore was obviously similar to predicting the S\&P500 or other financial
index values at a certain future time. Whence various techniques which were used
in econophysics, like the detrended fluctuation analysis, the multifractals, the
moving average crossing techniques, etc. could be attempted from scratch.

Beside the weather (temperature) derivatives other effects are of interest.
Climate is said to be fast changing nowadays.  Much is said and written about
e.g. the ozone layer and the Kyoto ''agreement''. The El Ni$\tilde{n}$o system is
a great challenge to scientists. Since there is some data available under the
form of time series, like  the Southern Oscillation Index, it is of interest to
look for trends, coherent structures, periods, correlations in noise, etc. in
order to bring some knowledge, if possible basic parameters, to this
meteorological field and expect to import some modern statistical physics ideas
into such climatological phenomena.

It appeared that other data are  available like those obtained under various
experiments,   put into force by various agencies, like   the Atlantic
Stratocumulus Transition Experiment (ASTEX) for ocean surfaces or those of the
Atmospheric Radiation Measurement Program (ARM) of the US Department of Energy,
among others. Much data is about cloud structure, e.g the cloud base height
evolution, on liquid water paths, brightness temperature, ...; they can often
freely downloaded  from the web. Therefore many time series can be analyzed.

However it appeared that the  data is sometimes of rather limited value because
of the lack of precision, or are biased because the raw data is already
transformed through models, and arbitrarily averaged (''filtered'') whence even
sometimes lacking the meaning it should contain. Therefore a great challenge
comes through in order to sort out the wheat from the chaff in order to develop
meaningful studies.

In Sect.2, I will  comment on the history of meteorology, observe that  the
evolution of such an old science is slow and limited by various $a$ $priori$
factors. Some basic recall on clouds and their role on climate and weather will
be made (Sect.3). This should remind us that the first modern ideas of
statistical physics were implemented on cloud studies through fractal $geometry$.
Indeed, modern and pioneering work on clouds  is due to Lovejoy who looked at the
perimeter-area relationship of rain and cloud areas \cite{lovejoy1}, fractal
dimension of their shape or ground projection. He discovered the statistical
self-similarity of cloud boundaries through area-perimeter analyses of the
geometry from satellite pictures. He found the fractal dimension $D_p  \simeq  4/3$ over a spectrum of 4
orders of magnitude in size, for small fair weather cumuli ($\sim$ 1021 km) up to
huge stratus fields ($\sim$ 103 km). Occasional scale breaks have been reported
\cite{15,17} due to variations in  cloudiness. Cloud size distributions have also
been studied from a scaling point of view. It is hard to say whether there is
perfect scaling\cite{Cahalan82,neggers,rodts}; why should there be scaling ?

I will point out, as others do, the basic well known pioneering work of modern
essence, like the Lorenz model. It was conceived in order to induce
predictability, but it turned out rather to be the basic nonlinear dynamical
system describing chaotic behavior.  However this allows for bringing up to its
level the notion of fractal ideas for meteorology work, thus scaling laws, and
modern data analysis techniques. I will recall most of the work to which I have
contributed,  being aware that I am failing to acknowledge many more  important
reports than those, - for what I deeply apologize.

There is a quite positive view of mine however. Even though the techniques have
not yet brought up many codes implemented in weather and climate evolution
prediction,   it was recently stressed \cite{nature1} in a sarcastic way (''Chaos
: useful at last?'') that  some applications of nonlinear dynamics  ideas  are
finding their way onto weather prediction\cite{6}, even though it  has to be said
that there is much earlier work on the subject\cite{7}.

There are  (also) very interesting lecture notes on the web for basic modules on
meteorological training courses, e.g. one available through ECMWF
website\cite{ECMWF}. But I consider that beyond the scientifically sound and
highly sophisticated computer models, there is still space for simple technical
and useful approaches, based on standard statistical physics techniques and
ideas, in particular based on the scaling hypothesis for phase transitions
\cite{StanleyPTbook} and percolation theory features \cite{StaufferP2book}. These
constraint allow  me to shorten the reference list ! A few examples will be found
in Sect. 4.

At the end of this introduction, I would crown the paper with references to two
outstanding scientists. First let me recall Friedmann\footnote{A. Friedmann
(1888-1925)  had a bad experience with his ideas, e.g. rejected by Einstein.
''Thus'' in July 1925 Friedmann made a record-breaking ascent in a balloon to
7400 meters to make meteorological and medical observations. Near the end of
August 1925 Friedmann began to feel unwell. He was diagnosed as having typhoid
and taken to hospital where he died two weeks later. He never became a
sociologist.} who said that ''if you can't be a good mathematician, you try to
become a good physicist, and those who can't become meteorologists''. Another,
Heisenberg was surely aware about errors and prediction difficulties resulting
from models.\footnote{W. Heisenberg (1901-1976) had much difficulty to get his
Ph.D. at TU Munchen because he was not  a good experimentalist!} Both men should
be guiding us to new endeavors with modesty anyway.

\section{ HISTORICAL INTRODUCTION}

From the beginning of times, the earth, sky, weather have been of great concern.
As soon as agriculture, commerce, travelling on land and sea prevailed, men have
wished to predict the weather. Later on airborne machines need atmosphere
knowledge and weather predictions for best flying. Nowadays there is much money
spent on weather predictions for sport activities. It is known how the knowledge
of  weather (temperature, wind, humidity, ..) is relevant, (even $fundamental$
!), e.g.  in sailing races or in Formula 1 and car rally races. Let it be
recalled the importance of knowing and predicting the wind (strength and
directions), pressure and temperature at high altitude for the (recent) no-stop
balloon round the world trip.

A long time ago,  druids and other priests were the up-to-date meteorologists. It
is known that many proverbs on weather derive from farmer observations, - one of
the most precise ones  reads (in french) ''Apr\`es la pluie, le beau temps'',
which is (still) correct, in spite of the Heisenberg principle, and modern
scientific advances.

After land travel and commerce, the control of the seas was of great importance
for economic, whence political reasons. Therefore there is no surprise in the
fact that at the time of  a British Empire, and the Dutch-Spanish-Portuguese
rivalry the first to draw sea wind maps was Halley\cite{bookmap}. That followed
the  ''classical'' isobaths and isoheights (these are geometrical measures !!!)
for sailors needing to go through channels. Halley, having also invented the
isogons (lines of equal magnetic fields) drew in ca. 1701, the first trade wind
and monsoon maps\cite{bookmap}, over the seas\footnote{Halley is also known for
the discovery of a comet bearing his name  and for using Breslau mortality
tables, - the basis of useful statistical work in actuary science.}. It may be
pointed out that he did not know Coriolis forces yet.

A second major step for meteorology seems to be due to Karl Theodor who between
1781 and 1792 was  responsible for the Palatinate Meteorological Society. He
invited (39) friends around the world\cite{bookmap}, from Massachusetts to Ural,
to make three measurements per day, and report them to him, in order to later
publish ''Ephemerides'', in Manheim.

I am very pleased to point out that Heinrich Wilhelm Brandes(1777-1834),
Professor of Mathematics and Physics at the University of Breslau was the
first\cite{bookmap} who had the idea of displaying weather data (temperature, air
pressure, a.s.o.) on geographical maps\footnote{It seems that H.W. Brandes left
Breslau to get his Ph.D. thesis in Heidelberg in 1826. Alas it seems that the
original drawings  are not available at this time. Where are they?}.

Later  von Humboldt (1769-1859)  had the idea to connect points in order to draw
isotherms\cite{bookmap}.  No need to say that this was a bold step, - the first
truly predictive step implying quantitative thermodynamics data. Most likely a
quite incorrect result. It is well known nowadays that various algorithms will
give various isotherms, starting from the same temperature data and coordinate
table. In the same line of lack of precision, there is no proof at all that the
highest temperature during the 2003 summer was 42.6$^{\circ}$C at the point of
measurement at Cordoba Airport on Aug. 12, 2003. There is no proof that the
highest temperature was 24.9$^{\circ}$C in Poland, in Warsaw-Ok\c{e}cie, on that
day in 2003. There is no proof that we will ever know the lowest temperature in
Poland in 2003 either. In fact the maximum or minimum temperature as defined in
meteorology\cite{meteoTdef,Huschke} are far from the ones acceptable in physics
laboratories. Therefore what isotherms are drawn from such data? They connect
data points which values are obtained at different times! What is their physical
meaning ? One might accept to consider that isotherms result from some truly
averaged temperature during one day (!) at some location (?). What is an average
temperature? In meteorology, it is NOT the ratio of the integral of the
temperature distribution function over a time interval to that time
interval\cite{meteoTdef,Huschke}. Nor is it a spatial mean. In fact, what is a
''mean temperature'' ...  for a city, a country, ... for the world? One might
propose that one has to measure the temperature everywhere and continuously in
time, then make an average. Questions are not only how many thermometers does one
need, but also what precision is needed. Does one need to distribute the
thermometers homogeneously?  What about local peculiarities? like nuclear plants
? Should we need a fractal distribution in space of thermometers? Might one use a
Monte Carlo approach to locate them such that statistical theories give some idea
on  error bars. What error bars ? There is no error bar ever given on weather
maps, in newspapers or TV, on radio and rarely in scientific publications. Errors
are bad ! and forgotten. There is rarely a certitude  (or risk) coefficient which
is mentioned. It might not be necessary for the public, but yet we know, and it
will be recalled later, that for computer work and predictions the initial values
should be well defined. Therefore it seems essential to concentrate on predicting
the uncertainty in forecast models of weather and climate as emphasized
elsewhere\cite{Palmer,Philander}.

\section{ CLIMATE and WEATHER. The role of clouds}

Up to von Humboldt there was  no correlation discussed,  no model of weather,
except for qualitative considerations, only through the influence of the earth
rotation, moon phases, Saturn, or Venus or Jupiter or constellation  locations,
etc. However the variables of interest were becoming to be known, but predictive
meteorology and more generally climate (description and) forecasting had still a
need for better observational techniques, data collecting, subsequent
analysis\cite{lovejoy2}, and model outputs.

Earth's climate is clearly determined by complex interactions between sun,
oceans, atmosphere, land and biosphere \cite{atmosphere,andrews}. The composition
of the atmosphere is particularly important because certain gases, including
water vapor, carbon dioxide, etc., absorb heat radiated from Earth's surface. As
the atmosphere warms up, it in turn radiates heat back to the surface that
increases the earth's ''mean surface temperature'', by some 30~K above the value
that would occur in the absence of a radiation-trapping atmosphere
\cite{andrews}. Note that perturbations in the concentration of the radiation
active gases do alter the intensity of this effect on the earth's climate.

Understanding the processes and properties that effect atmospheric radiation and,
in particular, the influence of clouds and the role of cloud radiative feedback
are issues of great scientific interest\cite{clouds,wielicki}. This leads to
efforts to improve not only models of the earth's climate but also predictions of
climate change\cite{Hasselmann}, as understood over long time intervals, in
contrast to shorter time scales for weather forecast. In fact, with respect to
climatology the situation is very complicated because one does not even know what
the evolution equations are. Since controlled experiments cannot be performed on
the climate system, one relies on using ad hoc models to identify
cause-and-effect relationships. Nowadays there are several climate models
belonging to many different centers\cite{climatemodelgroups}. Their web sites not
only carry sometimes the model output used to make images but also provide the
source code. It seems relevant to point out here that the stochastic resonance
idea was proposed to describe climatology evolution\cite{Benzi}.

Phenomena of interest occurring on short (!) time and small (!) space scales,
whence the weather, are represented through atmospheric models with a set of
nonlinear differential equations based on Navier-Stokes equations
\cite{LandauLifshitz} for describing fluid motion, in terms of mass, pressure,
temperature, humidity, velocity, energy exchange, including solar radiation ...
whence for predicting the weather\footnote{It is fair to mention Sorel 
work\cite{Sorel} about general motion of the atmosphere in order to explain
strong winds on the Mediterranean sea. The quoted reference
has some interesting introduction about previous
work}. It should be remembered that solutions of such
equations forcefully depend  on the initial conditions, and steps of
integrations. Therefore a great precision on  the temperature, wind velocity,
etc. cannot be expected and the solution(s) are only looking like a mess after a
few numerical steps\cite{Pasini}. The Monte Carlo technique suggests to introduce
successively a set of initial conditions, perform the integration of the
differential equations and make an average thereafter\cite{Pasini}. If  some
weather map is needed, a grid is used with constraints on the nodes, but
obviously the precision (!) is not remarkable, - but who needs it ?

It is hereby time to mention Lorenz's\cite{Lorenz} famous pioneering work who
simplified Navier-Stokes equations \footnote{Beautiful and thought provoking
illustrations can be found on various websites\cite{Lorenzstrangeattractors},
demonstrating Poincar\'e cross sections, strange attractors, cycles,
bifurcations, and the like.}. However, predicting the result of a complex
nonlinear interactions taking place in an open system is a difficult
task\cite{ramsey}.

Much attention has been paid recently \cite{physworld1,physworld2} to the
importance of the main components  of the atmosphere, in particular clouds,  (see
Appendix A) in the water three forms --- vapor, liquid and solid, for buffering
the global temperature against reduced or increased solar heating \cite{ou}.
Owing to its special properties, it is believed, that water establishes lower and
upper boundaries on how far the temperature can drift from today's. However, the
role of clouds and water vapor in climate change is not well understood. In fact,
there may be positive feedback between water vapor and other greenhouse gases.
Studies suggest that the heliosphere influences the Earth climate via  mechanisms
that affecting the cloud cover \cite{marsh,svens}. Surprisingly the influence of
solar variability is found to be strong on low clouds (3 km), whence pointing to
a microphysical mechanism involving aerosol formation enhanced by ionization due
to cosmic rays.

At time scales of less than one day, significant fluxes of  heat, water vapor and
momentum are exchanged due to entrainment, radiative transfer, and/or
turbulence\cite{andrews,16,19,116,20,119}. The turbulent character of the motion
in the atmospheric boundary layer (ABL) is one of its most important features.
Turbulence can be caused by a variety of processes, like thermal convection, or
mechanically generated by wind shear, or following interactions influenced by the
rotation of the Earth\cite{19,20}. This complexity of physical processes and
interactions between them create a variety of atmospheric formations. In
particular, in a cloudy ABL the radiative fluxes produce local sources of heating
or cooling within the mixed-layer and therefore can greatly influence its
turbulent structure and dynamics, especially in the cloud base.

The atmospheric boundary layer is defined by its inner (surface)
layer\cite{atmosphere,andrews,16,19}. In an unstably stratified ABL the
dominating convective motions are generated by strong surface heating from the
Sun or by cloud-topped radiative cooling processes\cite{andrews}.  In contrast, a
stably stratified ABL occurs mostly at night in response to the surface cooling
due to long-wave length radiation emitted into the space.

In presence of clouds (shallow cumulus, stratocumulus or stratus) the structure
of the ABL is modified because of the radiative fluxes. Thermodynamical phase
changes become important. During cloudy conditions one can distinguish mainly :
(i) the case in which the cloud and the sub-cloud layers are fully coupled; (ii)
two or more cloud layers beneath the inversion, with the lower layer well-mixed
with an upper elevated layer, decoupled from the surface mixed layer or (iii) a
radiation driven elevated mixed cloud layer, decoupled from the surface.

Two practical cases can be considered : the marine ABL and the continental ABL.
The former is characterized by a high concentration of moisture. It is wet,
mobile and has a well expressed lower boundary. The competition between the
processes of radiative cooling, entrainment of warm and dry air from above the
cloud and turbulent buoyancy fluxes determine the state of equilibrium of the
cloud-topped marine boundary layer\cite{andrews}. The continental ABL is usually
dryer, less mobile, better defined lower and upper boundaries. Both cases have
been investigated for their scaling properties\cite{nadia,DMWC96,MDWC97}

\section{ Modern Statistical Physics Approaches}

The modern paradigm in statistical physics is that systems obey ''universal''
laws due to the underlying nonlinear dynamics independently of microscopic
details. Therefore it can be searched in meteorology whether one can obtain
characteristic quantities using the modern statistical physics methods as done in
other laboratory or computer investigations. To distinguish cases and patterns
due to ''external field'' influences or mere self-organized situations in
geophysics phenomena\cite{Turcotte} is not obvious indeed. What sort of feedback
can be found? or neglected ? Is the equivalent of chicken-egg priority problem in
geophysics easily solved? The coupling between human activities and deterministic
physics is hard to model on simple terms\cite{Corti}, or can even be
rejected\cite{Karner}.

Due to the nonlinear physics laws governing the phenomena in the atmosphere, the
time series of the atmospheric quantities are usually
non-stationary\cite{Karner,Davis} as revealed by Fourier spectral analysis, -
whih is  usually the first technique to use. Recently, new techniques have been
developed that can systematically eliminate trends and cycles in the data and
thus reveal intrinsic dynamical properties such as correlations that are very
often masked by nonstationarities,\cite{r14,BA}. Whence many studies  reveal
long-range power-law correlations in geophysics time
series\cite{Turcotte,Davis,FraedrichBlender} in particular in
meteorology\cite{bunde1,bunde2,BTpaper,Tsonisetal1,Tsonisetal2,TalknerWeber,buda,40a}.
Multi-affine properties
\cite{MDWC97,wavelets,DM94,Davis2,genpol,66b,klono,kita,turbwavelet,mfatmoturb}
can also be identified, using singular spectrum or/and wavelets.

There are different levels of essential interest for sorting out correlations
from data, in order to increase the confidence in
predictability\cite{MalamudTurcotte}. There are investigations based on long-,
medium-, and short-range horizons. The $i$-diagram variability ($iVD$) method
allows to sort out some short range correlations. The technique has been used on
a liquid water cloud content data set taken from the Atlantic Stratocumulus
Transition Experiment (ASTEX) 92 field program \cite{astexiVD}.  It has also been
shown that the random matrix approach can be applied to the empirical correlation
matrices obtained from the analysis of the basic atmospheric parameters that
characterize the state of atmosphere\cite{RMTStatAtmCorr}. The principal
component analysis technique is a standard technique\cite{PCA} in meteorology and
climate studies. The Fokker-Planck equation for describing the liquid water path
\cite{kijats}  is also of interest. See also some tentative search for power law
correlations in the Southern Oscillation Index fluctuations characterizing El
Ni$\tilde{n}$o\cite{prenino}. But there are many other works of
interest\cite{68b}.

\subsection{Ice in cirrus clouds}

In clouds, ice appears in a variety of forms, shapes, depending on the formation
mechanism and the atmospheric conditions\cite{15,17,20,wavelets,18,210} The cloud
inner structure, content, temperature, life time, .. can be studied. In cirrus
clouds, at temperatures colder than about $-40^{\circ}$~C ice crystals form.
Because of the vertical extent, ca. from about 4 to 14~km and higher, and the
layered structure of such clouds one way of obtaining some information about
their properties is mainly by using ground-based remote sensing instruments (see
Appendix B), and searching for the statistical properties (and correlations) of
the radio wave signal backscattered from the ice crystals. This backscattered
signal received at the radar receiver antenna is known to depend on the ice mass
content and the particle size distribution. Because of the vertical structure of
the cirrus cloud it is of interest to examine the time correlations in the
scattered signal on the horizontal boundaries, i.e., the top and bottom, and at
several levels within the cloud.

We have reported\cite{40a}  along the DFA  correlations in the fluctuations of
radar signals obtained at isodepths of $winter$ and $fall$ cirrus clouds. In
particular we have focussed attention  on three quantities: (i) the
backscattering cross-section, (ii) the Doppler velocity and (iii) the Doppler
spectral width. They correspond to the physical coefficients used in Navier
Stokes equations to describe flows, i.e. bulk modulus, viscosity, and thermal
conductivity. It was found that power-law time correlations exist with a
crossover between regimes at about 3~to 5~min, but also $1/f$ behavior,
characterizing the top and the bottom layers and the bulk of the clouds. The
underlying mechanisms for such correlations likely originate in ice nucleation
and crystal growth processes.

\subsection{Stratus clouds}

In another case, i.e. for stratus clouds,  long-range power-law correlations
\cite{BTpaper,buda} and multi-affine properties\cite{DMWC96,MDWC97,kita}   have
reported  for the liquid water fluctuations, beside the spectral
density\cite{Gerber}. Interestingly, stratus cloud data retrieved from the
radiance, recorded as brightness
temperature,\footnote{http://www.phys.unm.edu/~duric/phy423/l1/node3.html} at the
Southern Great Plains central facility and operated in the vertically pointing
mode\cite{errorref} (see Appendix B for a brief technical note on
instrumentation) indicated for the Fourier spectrum $S(f) ~\sim ~ f^{-\beta}$, a
$\beta$ exponent equal to $1.56\pm 0.03$ pointing to a nonstationary time series.
The DFA statistical method applied on the stratus cloud brightness microwave
recording\cite{BTpaper,JAM} indicates the existence of long-range power-law
correlations  over a two hour time.

Contrasts in behaviors, depending on seasons can be pointed out. The DFA analysis
of liquid water path data measured in April 1998  gives a scaling   exponent
$\alpha = 0.34 \pm 0.01$ holding from 3 to 60 minutes. This scaling range is
shorter than the 150~min scaling range\cite{BTpaper} for a stratus cloud in
January 1998 at the same site. For longer correlation times a crossover to
$\alpha=0.50 \pm 0.01$ is seen up to about 2 h, after which the statistics of the
DFA function is not reliable.

However a change in regime from Gaussian to non-Gaussian fluctuation regimes has
been clearly defined for the cloud structure changes using a finite size (time)
interval window. It has been shown that the DFA exponent turns from a low value
(about 0.3) to 0.5 before the cloud breaks. This indicates that the stability of
the cloud, represented by antipersistent fluctuations is (for some unknown reason
at this level)  turning into a system for which the fluctuations are similar to a
pure random walk. The same type of finding was observed for the so called Liquid
Water Path\footnote{The liquid water path (LWP) is the amount of liquid water in
a vertical column of the atmosphere; it is measured in cm$^{-3}$; sometimes in cm
!!!} of the cloud.

The value of $\alpha \approx 0.3$ can be interpreted as the $H_1$ parameter of
the multifractal analysis of liquid water content\cite{DMWC96,MDWC97,DM94} and of
liquid water path \cite{kita}. Whence, the appearance of broken clouds and clear
sky following a period of thick stratus can be interpreted as a non equilibrium
transition or a sort of fracture process in more conventional physics. The
existence of a crossover suggests two types of correlated events as in classical
fracture processes: nucleation and  growth of diluted droplets.  Such a marked
change in persistence implies that specific fluctuation correlation dynamics
should be usefully inserted as ingredients in {\it ad hoc} models.

The non equilibrium nature of the cloud structure and content\cite{seker} should
receive some further thought henceforth. It would have been interesting to have
other data on the cloud in order to understand the cause of the change in
behavior.

\subsection{Cloud base height}

The variations in the local $\alpha$-exponent (''multi-affinity'') suggest that
the nature of the correlations change with time, so called intermittency
phenomena. The evolution of the time series can be decomposed into successive
persistent and anti-persistent sequences. It should be noted that the
intermittency of a signal is related to existence of extreme events, thus a
distribution of events away from a Gaussian distribution, in the evolution of the
process that has generated the data. If the tails of the distribution function
follow a power law, then the scaling exponent defines the critical order value
after which the statistical moments of the signal diverge. Therefore it is of
interest to probe the distribution of the fluctuations of a time dependent signal
$y(t)$ prior investigating its intermittency.  Much work has been devoted to the
cloud base height\cite{genpol,66b,klono}, under various ABL conditions, and the
LWP\cite{kita,kijats}. Neither the distribution of the fluctuations of liquid
water path signals nor those of the cloud base height appear to be Gaussian. The
tails of the distribution follow a power law pointing to ''large events'' also
occurring in the meteorological (space and time) framework.  This may suggest
routes for other models.

\subsection{Sea Surface Temperature}

Time series analysis methods searching for power law exponents allow to look from
specific view points, like atmospheric\cite{Pelletier}  or sea surface
temperature fluctuations\cite{Monetti}. These are of importance for weighing
their impacts on regional climate, whence finally to greatly increase
predictability of precipitation during all seasons.

Currently, scientists rely on climate patterns derived from global sea surface
temperatures (SST) to forecast precipitation e.g.  the U.S. winter. For example,
rising warm moist air creates tropical storms during El Ni$\tilde{n}$o years, a
period of above average temperatures in the waters in the central and eastern
tropical Pacific. While the tropical Pacific largely dictates fall and winter
precipitation levels, the strength of the SST signal falls off by spring through
the summer. For that reason, summer climate predictions are very difficult to
make.

Recently we have attempted to observe whether the fluctuations in the Southern
Oscillation index ($SOI$) characterizing El Ni$\tilde{n}$o were also prone to a
power law analysis. For the  $SOI$ monthly averaged data time interval 1866-2000,
the tails of the cumulative distribution of the fluctuations of $SOI$ signal it
is found that large fluctuations are more likely to occur than the Gaussian
distribution would predict. An antipersistent type of correlations exist for a
time interval ranging from about 4 months to about 6 years. This leads to favor
specific physical models for El Ni$\tilde{n}$o description\cite{prenino}.

\section{Conclusions}

Modern statistical physics techniques for analyzing atmospheric time series
signals indicate scaling laws (exponents and ranges) for correlations. A few
examples have been given briefly here  above, mainly from contributed papers in
which the author has been involved.  Work by many other authors have not been
included for lack of space. This brief set of comments is only intended for
indicating how meteorology and climate problems can be tied to scaling laws and
inherent time series data analysis techniques.  Those ideas/theories have allowed
me to reduce the list of quoted references, though even like this I might have
been unfair. One example can be recalled in this conclusion to make the point:
the stratus clouds break when the molecule density fluctuations become Gaussian,
i.e. when the molecular motion becomes Brownian-like. This should lead to better
predictability on the cloud evolution. Many other examples can be imagined. In
fact, it would be of interest for predictability models to examine whether the
long range fluctuations belong to a Levy-like or Tsallis or ... rather than to a
Gaussian distribution as in many self organized criticality models. This answer,
if positive, would enormously extend the predictability range in weather
forecast.

{\bf Acknowledgments} \vskip 0.6cm

Part of this studies have  been supported through an Action Concert\'ee Program
of the University of Li$\grave e$ge (Convention 02/07-293). Comments by A.
P\c{e}kalski, N. Kitova, K. Ivanova and C. Collette are greatly appreciated.

\vskip 1cm {\bf Appendix A.  CLOUDS}

It may be of interest  for such a type of proceedings to define clouds, or at
least to review briefly cloud classifications. Clouds can be classified by their
altitude

\begin{enumerate} \item  High clouds: High clouds are those having a cloud base
above 6000 m, where there is little moisture in the air. Typically they contain
ice crystals, often appear thin and wispy, and sometimes appear to create a halo
around the sun or the moon. High clouds are Cirrus (Ci), Cirrostratus (Cs),
Cirrocumulus (Cc) and Cumulonimbus (Cb)

\item  Middle clouds: Middle clouds have a cloud base between 2000 m and 6000 m.
These clouds are usually composed of water droplets, but sometimes contain ice
crystals, if the air is cold enough. Middle clouds are Altostratus (As) and
Altocumulus (Ac).

\item  Low clouds: Low clouds have a cloud base at less than 2000 m and appear
mostly above the sea surface. Usually composed of water droplets except in winter
at high latitudes when surface air temperature is below freezing. Low clouds are
Stratus (St), Stratocumulus (Sc) and Cumulus (Cu).

\end{enumerate}

or by their shape:

\begin{enumerate} \item  Cirrus: The cirrus are thin, hair-like clouds found at
high altitudes.

\item  Stratus: The stratus clouds are layered clouds with distinguished top and
bottom. Found at all altitudes, they generally thinner when high: (i) High: The
high stratus are called Cirrostratus; (ii) Middle: The middle stratus are called
Altostratus; (iii) Low: The low ones are stratus and nimbostratus

\item  Cumulus: The cumulus clouds are fluffy heaps or puffs. They are found at
all altitudes; generally they are lighter and smaller when become high: (i) High:
The high cumulus are called Cirrocumulus; (ii) Middle: The middle cumulus are
called Altocumulus; (iii) Low: The low ones are Stratocumulus and Cumulus.

\item  Cumulonimbus: The cumulonimbus clouds are thunder clouds. Some large
cumulus clouds are height enough to reach across low, middle and even high
altitude.

Note that due to their relatively small sizes and life time cumulus clouds
produce short time series for  remote sensing measurements (see App. B).
Therefore such clouds and data series are not often suitable for many techniques
mentioned in this report. It is fair to point out on such clouds the study
pertaining to a phenomenon of Abelian nature, rain fall\cite{extreme,pinho}. Much
work has been devoted to rain of course, see e.g. Andrade et
al.\cite{pinho,andrade,miranda} or Lovejoy et al.\cite{tessier,73a,73b,73c}

\end{enumerate} \vskip 1cm

{\bf APPENDIX B. Experimental techniques and data acquisition}

Quantitative observations of the atmosphere are made in many different ways.
Experimental/observational techniques to study the atmosphere rely on physical
principles. One important type of observational techniques is that of {\it remote
sounding}, which depends on the detection of electromagnetic radiation emitted,
scattered or transmitted by the atmosphere. The instruments can be placed at
aircrafts, on balloons or on the ground. Remote-sounding techniques can be
divided into {\it passive} and {\it active} types. In passive remote sounding,
the radiation measured is of natural origin, for example the thermal radiation
emitted by the atmosphere, or solar radiation transmitted or scattered by the
atmosphere. Most space-born remote sounding methods are passive. In active remote
sounding, a transmitter, e.g. a radar, is used to direct pulses of radiation into
the atmosphere, where they are scattered by atmospheric molecules, aerosols or
inhomogeneities in the atmospheric structure. Some of the scattered radiation is
then detected by some receiver. Each of these techniques has its advantages and
disadvantages. Remote sounding from satellites can give near-global coverage, but
can provide only averaged values of the measured quantity over large regions, of
order of hundreds of kilometers in horizontal extent and several kilometers in
the vertical direction.  Ground-based radars can provide data with very high
vertical resolution (by measuring small differences in the time delays of the
return pulses), but only above the radar site.

For a presentation of remote sensing techniques the reader can consult many
authors\cite{andrews,errorref,73c,westwater78,westwater,rees}, or  the ARM
site\cite{arm,web,ceilometer}. For example, microwave radiometers work at
frequencies of 23.8 and 31.4 GHz. At the DOE ARM program SGP central facility, in
the vertically pointing mode, the radiometer makes sequential 1 s radiance
measurements in each of the two channels while pointing vertically upward into
the atmosphere. After collecting these radiances the radiometer mirror is rotated
to view a blackbody reference target. For each of the two channels the radiometer
records the radiance from the reference immediately followed by a measurement of
a combined radiance from the reference and a calibrated noise diode. This
measurement cycle is repeated once every 20 s. Note that clouds at 2 km of
altitude moving at 10 $\rm m \,s^{-1}$ take 15 s to advect through a radiometer
field-of-view of approximately 5$^{\circ}$. The 1 s sky radiance integration time
ensures that the retrieved quantities correspond to a specific column of cloud
above the instrument.

The  Belfort Laser Ceilometer (BLC)\cite{73c,ceilometer} detects clouds by
transmitting pulses of infrared light (l=910 nm) vertically into the atmosphere
(with a pulse repetition frequency fr=976.6 Hz) and analyzing the backscattered
signals from the atmosphere.   The ceilometer actively collects backscattered
photons for about 5 seconds within every 30-second measurement period. The BLC is
able to measure the base height of the lowest cloud from 15 up to 7350 m directly
above mean ground level. The ceilometer works with a 15 m spatial resolution and
reaches the maximum measurable height of 4 km. The time resolution of CBH records
is 30 seconds.

 \end{document}